\documentstyle[pra,aps]{revtex}
\begin{document}
\title{\large \bf Kinetics of Strongly Non-Equilibrium 
Bose-Einstein Condensation}
\author {Boris Svistunov}
\address{
Russian Research Center ``Kurchatov Institute", 123182 Moscow, Russia}

\maketitle
\begin{abstract}
We consider the ordering kinetics in a strongly non-equilibrium state of
a (weakly) interacting Bose gas, characterized, on one hand, by large 
occupation numbers, and, on the other hand, by the absence of long-range 
order. Up to higher-order corrections in inverse occupation numbers,
the evolution is described by non-linear Schr\"{o}dinger equation with
a turbulent initial state. The ordering process is rather rich and
involves a number of qualitatively different regimes that take place 
in different regions of energy space. Specially addressed is the case of
evolution in an external potential.
\vspace{0.5cm}
\end{abstract}

\section{Introduction}
Kinetics of Bose-Einstein condensation (BEC) in a weakly interacting Bose gas
is one of the most fundamental problems of non-equilibrium statistical mechanics. 
The exciting progress in the experiment with BEC in ultracold gases initiated 
by the pioneer works \cite{BEC} opens up an opportunity of laboratory study of 
non-trivial regimes of BEC kinetics. 

From the very beginning it should be realized that the statement of the
problem of BEC kinetics involves a number of aspects that
are of crucial importance to the very character of the evolution process.
The nature of the process strongly depends on how the BEC is being achieved:
Say, by slow cooling, or by self-evolution of an essentially non-equilibrium
initial state. More generally, it is important to take into account whether the 
considerable deviation from equilibrium occurs only at some sufficiently 
large length scales and only in the fluctuation region (so that the kinetics
is of universal character and does not reflect specifics of weakly
interacting system), or the non-equilibrium situation arises far enough from the
critical region and the resulting ordering kinetics is characteristic only to
weakly interacting gas. Obviously, the picture of evolution can be ``trivialized" 
if some portion of condensate is already present in the initial state. Finally,
and especially importantly for the realistic case of a trapped gas, finite size of
the system, or just only that of the condensate can partially or completely
change the relaxation scenario, if this size turns out to be less than some 
correlation length relevant to the ordering process in the infinite system.

In this paper we concentrate on a statement of BEC kinetics problem that we 
believe to be the most characteristic of the case of 
weakly interacting gas. Namely, we consider the self-evolution of weakly 
interacting gas with a strongly non-equilibrium initial state. To maximally
simplify the consideration without qualitatively changing the nature of the 
process, we assume that in the initial state all occupation numbers are
either much larger than unity, or equal to zero, and that there are no correlations
between different single-particle modes. The advantage of choosing such an initial
condition is that from the very beginning one can employ classical-field description
in terms of non-linear Schr\"{o}dinger equation (NLSE), which in the theory of Bose
gases is known as Gross-Pitaevskii equation \cite{GP}, 
with a certain turbulent initial condition. It should be stressed that, in contrast 
to a wide-spread prejudice, the very description in terms of NLSE does not imply
the presence of condensate, or any sort of dynamical phase transition (see, e.g., 
discussion in \cite{KS2}). The question of the presence of condensate, or,
more generally, the question of (long-range) order is the question of the 
(long-range) structure of corresponding classical field.

Hence, the dynamical model for our problem reads ($\hbar=1$)
\begin{equation}
i \frac{\partial \psi}{\partial t} \, = \, 
-\frac{\Delta}{2m} \psi + V(\vec{r})\psi + U \! \mid \psi \mid^2 \psi \; ,
\label{NLSE}
\end{equation}
where $\mid \psi \mid^2$ is interpreted as particle density 
(not the condensate density!), $m$ is the particle mass, $V(\vec{r})$ is the external potential;
$U=4 \pi a /m$ is the vertex of the effective pair interaction, $a$ is the scattering 
length. To introduce the initial condition to (\ref{NLSE}) one has to consider
the expansion of $\psi(\vec{r},t)$ in terms of eigen modes 
$\varphi_{\varepsilon}(\vec{r})$  
[$(-\Delta/2m+V)\varphi_{\varepsilon}=\varepsilon \varphi_{\varepsilon}$] of 
linear part of NLSE:
$\psi(\vec{r},t) = \sum_{\varepsilon} a_{\varepsilon}(t)
\varphi_{\varepsilon}(\vec{r})$. Then, at the initial moment $t=0$, 
the phases of the complex amplitudes
$a_{\varepsilon}$ can be considered as random, while $\mid a_{\varepsilon} \mid^2$
is identified with the occupation number $n_{\varepsilon}$ of the mode $\varepsilon$
(see, e.g., \cite{KS2} for more details). To the best of our knowledge, 
the first formulation of BEC kinetics problem in terms of NLSE was given in \cite{LY}. 

A full-scale numeric simulation of NLSE with the turbulent initial condition
could, in principle, cross almost all the t's in the strongly non-equilibrium BEC
kinetics problem. Such a simulation has not been done yet \cite{Damle}. Nevertheless,
we will see that from general considerations it is possible to propose the evolution 
scenario and to obtain all relevant estimates.

In very general terms, the direction in which the field $\psi$ will evolve 
is clear from the following considerations. First, it is natural to
expect that the system must relax to a certain equilibrium state. Secondly,
this equilibrium state should correspond to zero temperature, since the classical
field described by the equation (\ref{NLSE}) forms a heatbath at absolute zero
with respect to itself. Hence, if there is a stable groundstate (that is 
if $U > 0$) for a given particle density, then the system should approach it
in this or that way, the excess energy being carried away (to higher and higher
harmonics) by ever decreasing portion of high-frequency fraction of the field.
At the final stage of evolution all the particles are condensed except for an
infinitesimally small high-frequency portion. 

Though the general tendency of evolution is clear, the particular relaxation
scenario is not at all self-evident. A detailed analysis \cite{Sv,KSS,KS1} 
leads to a rather sophisticated scenario that involves a number of qualitatively 
different stages. 
The evolution starts with an explosion-like wave in energy space, propagating 
from higher energies towards the lower ones, that leads to a formation of a 
specific power-law distribution of particles. Immediately after its formation, 
this distribution starts to relax. Simultaneously, in the low-energy region the
so-called coherent regime sets in that leads to the formation of quasi-condensate 
correlation properties. Basically, the quasi-condensate state corresponds
to what is known in the theory of superfluidity as the state of superfluid
turbulence. It can be viewed as a condensate containing a tangle of vortex lines
(plus a specific sharply non-equilibrium distribution of long-wave phonons). 
The formation of the quasicondensate occurs very rapidly (characteristic time is
much smaller than the time of the wave formation). In contrast to it,
the final stage of long-range ordering, associated with relaxing superfluid turbulence
and long-wave phonons, takes a macroscopically large time.

In the present paper we render the homogeneous BEC scenario \cite{Sv,KSS,KS1} 
(sections \ref{kin} and \ref{coh}) and project it onto the case of a trapped gas.
We find out that in an external potential the evolution picture can be even 
more rich.

\section{Kinetic regime}
\label{kin}

During some initial period of evolution the correlations between different amplitudes
$a_{\varepsilon}$ are vanishingly small. Such a regime (known in the theory
of non-linear classical-field dynamics as weak turbulence \cite{Nazarenko}) 
admits a description in terms of kinetic equation. 
This stage is thus referred to as kinetic stage.

Kinetic equation corresponding to the weak-turbulence regime of non-linear Schr\"{o}dinger
equation  belongs to a generic class of scale-invariant models with four-wave 
particle- and energy-conserving interaction, that
allows an analysis of evolution kinetics in general terms (see, e.g., \cite{Sv}).
For the BEC kinetics the analysis suggests that there are two alternative ways of 
the initial evolution: (i) shrinking of the particle distribution as a whole towards
$\varepsilon=0$ (during infinite time), or (ii) a specific wave in the energetic space 
leading to a singularization of distribution at the point $\varepsilon=0$ at some 
finite time moment $t=t_*$. The answer to the question of which scenario takes place for a
given model depends only on the scaling properties of the collision integral
and the density of states; and the case of NLSE corresponds to the scenario (ii).

The evolution at the beginning of the kinetic 
stage results in the formation of self-similar wave in the energy space propagating in an 
explosion-like fashion from the high-energy region (where the particles are 
initially distributed) towards lower energy scales.
Corresponding self-similar solution of the kinetic equation has the form \cite{Sv}
\begin{equation}
n_{\varepsilon}(t) = A \varepsilon_0^{-\alpha}(t) 
f(\varepsilon/\varepsilon_0(t)) \;, \;\;\; t \leq t_* \; , 
\label{wave_a}
\end{equation}
\begin{equation}
\varepsilon_0(t) = B \mid t_* -t \mid^{1/2(\alpha-1)}\; . 
\label{wave_b}
\end{equation}
Here $A$ and $B$ are dimensional constants depending on the initial condition
and related to each other by the formula 
$B=\mathrm{const} (m^3 U^2 A^2)^{1/2(\alpha-1)}$.
The dimensionless function $f$ (numeric data for $f$ see in \cite{Sv}) is 
defined up to an obvious scaling freedom.  
The explosion character of the evolution guarantees that the wave reaches
the point $\varepsilon=0$ at some finite time moment $t=t_*$ 
[$t=0$ corresponds to the beginning of evolution], the value of $t_*$ 
being on the order of the typical time of (stimulated) collisions in the gas 
at $t=0$. Physically, this explosion-like evolution
is supported by the stimulation of the collision rate at the head of the wave,
$\varepsilon \sim \varepsilon_0$, by ever growing occupation numbers.

Generally speaking, the index $\alpha$ in (\ref{wave_a})-(\ref{wave_b}) 
cannot be established from the scaling
properties of the collision term of the kinetic equation, 
being related thus to the particular
form of the latter. It is possible, however, to specify lower and upper limits
for $\alpha$ following from the consistency of (\ref{wave_a}) and (\ref{wave_b}) 
with the requirement that these formulae describe an explosion-like singularization
of distribution (rather than infinite-time shrinking). To this end we note that 
from the scale invariance it follows that $f(x)$ behaves like some power of $x$
at $x \gg 1$. At $t=t_*$ the occupation numbers have to be finite at
$\varepsilon > 0$, hence
\begin{equation}
f(x) \to x^{-\alpha} \;\; \mathrm{at} \;\; x \to \infty \;  . 
\label{f}
\end{equation}
The requirement that the particle distribution does not shrink as a whole
implies that the number-of-particles integral for the distribution (\ref{wave_a}), 
(\ref{f}) is divergent at $\varepsilon \to \infty$. This immediately yields
$\alpha < 3/2$. The condition $\alpha > 1$ is necessary 
for $\varepsilon_0$ (\ref{wave_b}) to approach zero at $t=t_*$. 
So we have $1 < \alpha < 3/2)$. The most accurate up-to-date numeric 
analysis of $\alpha$ was performed in \cite{ST} with the result
$\alpha \approx 1.24$.

At $t>t_*$ kinetic description is still valid for not so small energies,
but to obtain an adequate sewing with the solution (\ref{wave_a})-(\ref{wave_b})
one has to explicitly introduce the (quasi)condensate, employing
the conservation of the total number of particles (see the discussion in \cite{Sv}).
The structure of the self-similar solution
\begin{equation}
n_{\varepsilon}(t) = A \varepsilon_0^{-\alpha}(t) 
\tilde{f}(\varepsilon/\varepsilon_0(t)) \;, \;\;\; 
\varepsilon > 0 \;, \;\;\; t \geq t_* 
\label{back}
\end{equation}
[$\tilde{f}(x) \to f(x) \;\;$ at $\;\; x \to \infty$]
corresponds to a back wave in the energy space, destroying the singular
distribution created by the wave (\ref{wave_a})-(\ref{wave_b}). The particles being
released during this destruction go directly to quasicondensate. For the quasicondensate
density $n_0$ we thus have
\begin{equation}
n_0(t) = {A \over 4 \pi^2}  (2m)^{3/2} \varepsilon_0^{3/2-\alpha}(t)
\int_0^{\infty} d x \sqrt{x} [x^{-\alpha} 
- \tilde{f}(x)] 
\propto (t-t_*)^{(3-2\alpha)/4(\alpha-1)}\; .
\label{n_0}
\end{equation}
As follows from general considerations and is supported by direct numeric
analysis \cite{Sv}, $\tilde{f}(x) \propto 1/x$ at $x \ll 1$, which means that
the back wave creates a quasi-equilibrium distribution at $\varepsilon \ll 
\varepsilon_0(t)$ [with infinite at $t=t_*$ and ever decreasing afterwards 
temperature $\propto \varepsilon_0^{-\alpha}(t)$].

To estimate the parameter $A$ for a given initial conditions one extrapolates
the solution (\ref{wave_a})-(\ref{wave_b}) to a region of energies 
$\sim \varepsilon_{\mathrm{init}}$, where the particles were initially concentrated
with typical occupation numbers $n_{\varepsilon_{\mathrm{init}}}$. This immediately
yields $A \sim n_{\varepsilon_{\mathrm{init}}} \varepsilon_{\mathrm{init}}^{\alpha}$.

\section{Coherent regime}
\label{coh}

Strictly speaking, evolution in a kinetic regime does not lead to the ordering. It is
seen from the fact that the description in terms of kinetic equation is associated 
with the random phase approximation (RPA) and thus valid only when the phases of
$a_{\varepsilon}$'s are practically uncorrelated. In such a state even
local order (quasicondensate) is absent. Quasicondensation implies a strong change
of the correlation properties as compared to the RPA state \cite{KSS}. 
It occurs in the regime of {\it strong} turbulence (so-called coherent regime), 
when typical time of evolution is comparable to the
time of oscillation of the phases of relevant $a_{\varepsilon}$'s. The essence of the 
process of the quasicondensate formation is the transformation of the strong turbulence
into the state known as superfluid turbulence.

The degrees of freedom associated with the quasicondensate are the same as in a 
genuine condensate. These are phonons and topological defects (vortex lines). 
That is quasicondensate can be viewed as a condensate with (i) a tangle of
vortex lines and (ii) strongly non-equilibrium distribution of long-wave
phonons implying strong fluctuations of the phase of the quasicondensate
part $\psi_0$ of the field $\psi$ at large distances.

Given the solution (\ref{wave_a})-(\ref{wave_b}) of the kinetic equation, 
one readily estimates where and when the coherent regime sets in, and what is 
the typical density of the quasicondensate upon its formation. The characteristic time of 
evolution at the energy scale $\varepsilon$ is the collision time
$\tau_{\mathrm{coll}}^{-1}(\varepsilon) \sim m^3(U \varepsilon n_{\varepsilon})^2$.
For $n_{\varepsilon} \sim A \varepsilon^{-\alpha}$ the RPA criterion 
$\tau_{\mathrm{coll}}(\varepsilon) \varepsilon \gg 1$ becomes invalid at
$\varepsilon \sim \varepsilon_{\mathrm{coh}}=(m^3U^2A^2)^{1/(2\alpha -1)}$.
The distribution $n_{\varepsilon} \sim A \varepsilon^{-\alpha}$ at the
scale $\varepsilon_{\mathrm{coh}}$ is formed at the time $t_{\mathrm{coh}}$
obeing an obvious relation $\varepsilon_0(t=t_* -t_{\mathrm{coh}}) 
\sim \varepsilon_{\mathrm{coh}}$. Hence, $t_{\mathrm{coh}}$ estimates the time
moment when the coherent regime sets in. The time interval $\mid t_* -t_{\mathrm{coh}} 
\mid \varepsilon_{\mathrm{coh}}^{-1} \ll t_*$ is a typical time of the process
of quasicondensate formation (in the strong turbulent regime all characteristic times
and distances scale with $\varepsilon_{\mathrm{coh}}^{-1}$ and
$(m \varepsilon_{\mathrm{coh}})^{-1/2}$, correspondingly). The initial quasicondensate
density $n_0^{\mathrm{init}}$ is the density corresponding to the harmonics
$\varepsilon \sim \varepsilon_{\mathrm{coh}}$. It obeys an obvious 
relation $n_0^{\mathrm{init}}U \sim \varepsilon_{\mathrm{coh}}$, in accordance with 
the fact that in strong turbulent regime kinetic and potential energies are 
of the same order. The initial spacing between the vortex lines in the quasicondensate
scales as $(m \varepsilon_{\mathrm{coh}})^{-1/2}$.

The coherent regime and the back wave are practically independent processes, 
with a reservation that the coherent part of the field is being pumped with 
particles from the high-energy region. This pumping, however,
does not affect the character of the coherent evolution.

Relaxation of the quasicondensate towards genuine condensate goes in two directions: 
(i) relaxation of the vortex tangle [that is relaxation of the superfluid turbulence]
and (ii) relaxation of the long-wave phonons. Both processes require a macroscopically
large time (an analysis of this stage of evolution see in \cite{KS1}).

\section {External potential}
\label{pot}

In the experiments with trapped ultracold gases normally there takes place
the Knudsen regime, when free path length of a particle with the energy $\varepsilon$,
$l_{\mathrm{free}}(\varepsilon)$, is much larger than the typical radius of the 
particle's trajectory, $R_{\varepsilon}$. For definiteness we consider a parabolic
trap with all the three frequencies of the same order $\omega_0$. So that
$R_{\varepsilon} \sim \omega_0 v(\varepsilon)$ [$v(\varepsilon)$ is the 
typical velocity corresponding to the energy $\varepsilon$], and the condition 
$l_{\mathrm{free}}(\varepsilon) \gg R_{\varepsilon}$ is equivalent to
$\tau_{\mathrm{coll}}(\varepsilon) \omega_0 \gg 1$.

Knudsen regime is a very convenient starting point for analyzing kinetics 
in a potential. Almost in all qualitative aspects it corresponds to an
isotropic homogeneous case, since the distribution of particles depends
only on the two variables, $\varepsilon$ and $t$ (ergodic approximation). 
The main quantitative difference comes from the difference in the 
density of states, the scaling of the collision time remaining the same.

Initial picture of evolution is described by the self-similar wave 
(\ref{wave_a})-(\ref{wave_b})
with $\alpha \approx 1.6$ \cite{Gardiner,Stoof}.  The question then is:
What happens when $t \to t_*$? The effect of the potential can be
associated with two rather different reasons. The first reason is the essential 
discreteness of the low-lying energy levels, which drastically changes the kinetics when 
$\tau_{\mathrm{coll}}^{-1}(\varepsilon) < \Delta (\varepsilon)$, where
$\Delta (\varepsilon)$ is the typical interlevel spacing. The second reason is
the violation of the Knudsen regime at the head of the wave at some stage
of evolution [because of the decreasing $\tau_{\mathrm{coll}}(\varepsilon_0(t))$
with $t$].

Remarkably, the above-mentioned two circumstances arise always separately,
and (apart a certain cross-over region) there is nothing in between. More specifically,
as it immediately follows from the non-equality 
$\Delta (\varepsilon) \leq \omega_0$ and the fact that the relevant collision time
$\tau_{\mathrm{coll}}(\varepsilon_0(t))$ permanently decreases, if there occurs a 
break-down of Knudsen regime (in corresponding region of coordinate space
with a typical size $R_{\mathrm{Kn}}$ around the center of the potential), 
the discreteness of levels will never become relevant. 

The case, when the discreteness of levels starts to act within the Knudsen 
regime, is rather transparent physically and is studied to a large
extent both experimentally \cite{Miesner} and 
theoretically \cite{Gardiner,Stoof}. The evolution scenario in this case is as follows. 
When the wave reaches the scale where the level discreteness becomes relevant, its 
further propagation is suppressed (essentially discrete harmonics practically do not 
interact with each other) and the back wave is formed. At $\varepsilon < \varepsilon_0(t)$, 
the back wave generates quasi-equilibrium 
distribution with time-dependent permanently decreasing temperature $T(t)$ and permanently
increasing number of particles. Condensation thus occurs in a quasi-equilibrium
way, without the coherent stage (interaction between low-lying harmonics is
negligible), and the whole process can be described within the kinetic approach 
\cite{Gardiner,Stoof}. 

We are mostly interested in the case, when at some $t=t_{\mathrm{Kn}}$ the Knudsen 
regime breaks down for energies 
$\varepsilon_{\mathrm{Kn}} \sim \varepsilon_0(t=t_{\mathrm{Kn}})$.
From (\ref{wave_a})-(\ref{wave_b}) we estimate $\varepsilon_{\mathrm{Kn}} 
\sim [m^3U^2A^2/\omega_0]^{1/2(\alpha-1)}$. The size of corresponding spatial 
region is defined by $m \omega_0^2 R_{\mathrm{Kn}}^2 \sim 
\varepsilon_{\mathrm{Kn}}$. We argue that within the region 
$r < R_{\mathrm{Kn}}$ at $t>t_{\mathrm{Kn}}$ the external potential becomes 
{\it irrelevant} at least until the quasicondensate is formed, so that the most 
important evolution stage basically does not differ from the homogeneous case. 
Indeed, it is quite natural that further evolution within the region 
$r<R_{\mathrm{Kn}}$ will result
in the formation of {\it anti}-Knudsen regime $l_{\mathrm{free}}(\varepsilon)
 \ll R_{\mathrm{Kn}}$ for $\varepsilon \ll \varepsilon_0(t_{\mathrm{Kn}})$, 
because of increasing
collision rate with increasing the occupations numbers. In the anti-Knudsen
regime the evolution during the time period on the order of collision time
is insensitive to the external potential (the criteria for the Knudsen regime and for
the sensitivity to the potential within the collision time coincide). But this time is
enough to form the wave (\ref{wave_a})-(\ref{wave_b}) in the energy space 
(with the exponent $\alpha$ corresponding to the homogeneous case) 
and then to form quasicondensate. During the wave evolution in the energy
space, the free-path length of the particles with $\varepsilon \sim \varepsilon_0(t)$
is getting progressively smaller, which renders the proposed scenario self-consistent.
A minor deviation from the pure homogeneous picture is that now the moment $t_*$
depends on the distance from the center of the potential, so that the coherent
regime first should start at $r=0$ (the point of maximal initial density) and then 
gradually occupy all the anti-Knudsen region up to $r \sim R_{\mathrm{Kn}}$. 
By this moment the quasicondensate is formed at $r \leq R_{\mathrm{Kn}}$.
In terms of the total number of particles, $N$, and the typical single-particle 
energy of the initial distribution in the potential, $\varepsilon_{\mathrm{in}}$, 
the estimate for $R_{\mathrm{Kn}}$ is:
$R_{\mathrm{Kn}} \sim  [U^2 N^2 m^{5-2\alpha} \omega_0^{9-4\alpha} 
\varepsilon_{\mathrm{in}}^{2(\alpha -3)}]^{1/4(\alpha -1)}$.

On the basis of the above discussion one can introduce the parameter
\begin{equation}
p=N^2 \omega_0 m^3 U^2 (\omega_0 / \varepsilon_{\mathrm{in}})^{6-2\alpha} \; ,
\label{p}
\end{equation}
that determines which of the two regimes takes place under given initial conditions:
the discrete-harmonic regime ($p \ll 1$), or the superfluid-turbulence one
($p \gg 1$).

For more details on the ordering kinetics in a trapped gas see \cite{Sv_2}.

%

\end{document}